\documentclass[sigconf,natbib=true]{acmart}
\AtBeginDocument{%
  }

\setcopyright{acmlicensed}
\copyrightyear{2018}
\acmYear{2018}
\acmDOI{XXXXXXX.XXXXXXX}

\acmConference[Conference acronym 'XX]{Make sure to enter the correct
	conference title from your rights confirmation emai}{June 03--05,
	2018}{Woodstock, NY}
\acmISBN{978-1-4503-XXXX-X/18/06}

\usepackage{multirow}
\usepackage{colortbl}
\usepackage{subfigure}
\usepackage{tcolorbox}
\usepackage{acronym}
\usepackage{enumitem}
\acrodef{MLLMRec}{preference reasoning paradigm with graph refinement}
\acrodef{MMRS}{multimodal recommender systems}
\acrodef{GCNs}{graph convolutional networks}
\acrodef{BPR}{Bayesian Personalized Ranking}
\acrodef{MLPs}{multilayer perceptrons}
\acrodef{LLMs}{large language models}
\acrodef{MLLMs}{multimodal large language models}
\acrodef{CF}{collaborative filtering}
\acrodef{SSL}{self-supervised learning}
\newcommand{\done}[1]{\textcolor{black}{{#1}}}

\begin{document}

%%
%% The "title" command has an optional parameter,
%% allowing the author to define a "short title" to be used in page headers.
\title{MLLMRec: A Preference Reasoning Paradigm with Graph Refinement for Multimodal Recommendation}
%Listen to the Behavior: Synergizing Preference Reasoning and Graph Refinement for Multimodal Recommendation
%Listen to the Behavior: MLLM-driven Preference Reasoning for Multimodal Recommendation with Graph Refinement
%Read the Room
%MLLMRec: Exploring the Potential of Multimodal Large Language Models in Recommender Systems
%MLLMRec: User Preference Reasoning and Graph Refinement for Multimodal Recommendation
%MLLM-Driven Preference Reasoning Paradigm with Graph Refinement for Multimodal Recommendation
%Preference Matters! A MLLM-Driven User Preference Reasoner for Multimodal Recommendation
%Description Matters! MLLMRec: User Preference Reasoning for Multimodal Recommendation
%Language Reveals Profile: Exposing User Preference with Multimodal Large Language Model in Multimodal Recommendation
%MLLM as Preference Summarizer: Bridging Multimodal Signals to User Preference for Recommendation
%Preference Reasoning: Exposing User Preference with Multimodal Large Language Model in Multimodal Recommendation
%%
%% The "author" command and its associated commands are used to define
%% the authors and their affiliations.
%% Of note is the shared affiliation of the first two authors, and the
%% "authornote" and "authornotemark" commands
%% used to denote shared contribution to the research.

%\authornote{Both authors contributed equally to this research.}
%\authornotemark[1]
%\email{dangyuzhuo@nudt.edu.cn}
%\orcid{1234-5678-9012}

\author{Yuzhuo Dang}
\affiliation{%National Key Laboratory of Information Systems Engineering, National University of Defense Technology
  \institution{College of Systems Engineering, National University of Defense Technology}
  \city{Changsha}
  \country{China}
}
\email{dangyuzhuo@nudt.edu.cn}

\author{Xin Zhang}
\affiliation{%
	\institution{College of Systems Engineering, National University of Defense Technology}
	\city{Changsha}
	\country{China}
}
\email{zhangxin16@nudt.edu.cn}

\author{Zhiqiang Pan}
\affiliation{%
	\institution{College of Systems Engineering, National University of Defense Technology}
	\city{Changsha}
	\country{China}
}
\email{panzhiqiang@nudt.edu.cn}

\author{Yuxiao Duan}
\affiliation{%
	\institution{College of Systems Engineering, National University of Defense Technology}
	\city{Changsha}
	\country{China}
}
\email{duanyuxiao19@nudt.edu.cn}

\author{Wanyu Chen}
\affiliation{%
	\institution{College of Electronic Countermeasures, National University of Defense Technology}
	\city{Hefei}
	\country{China}
}
\email{wanyuchen@nudt.edu.cn}

\author{Fei Cai}
\authornote{Corresponding authors.}
\affiliation{%
	\institution{College of Systems Engineering, National University of Defense Technology}
	\city{Changsha}
	\country{China}
}
\email{caifei08@nudt.edu.cn}

\author{Honghui Chen}
\authornotemark[1]
\affiliation{%
	\institution{College of Systems Engineering, National University of Defense Technology}
	\city{Changsha}
	\country{China}
}
\email{chenhonghui@nudt.edu.cn}

%%
%% By default, the full list of authors will be used in the page
%% headers. Often, this list is too long, and will overlap
%% other information printed in the page headers. This command allows
%% the author to define a more concise list
%% of authors' names for this purpose.
\renewcommand{\shortauthors}{Dang et al.}

%%
%% The abstract is a short summary of the work to be presented in the
%% article.
\begin{abstract}
Multimodal recommendation combines the user historical behaviors with the modal features of items to capture the tangible user preferences, presenting superior performance compared to the conventional ID-based recommender systems.
However, existing methods still encounter two key problems in the representation learning of users and items, respectively:
(1) the initialization of multimodal user representations is either agnostic to historical behaviors or contaminated by irrelevant modal noise,
and (2) the widely used $K$NN-based item-item graph contains noisy edges with low similarities and lacks audience co-occurrence relationships.
To address such issues, we propose \acs{MLLMRec}, a novel preference reasoning paradigm with graph refinement for multimodal recommendation.
Specifically, on the one hand, the item images are first converted into high-quality semantic descriptions using a multimodal large language model (MLLM), thereby bridging the semantic gap between visual and textual modalities.
Then, we construct a behavioral description list for each user and feed it into the MLLM to reason about the purified user preference profiles that contain the latent interaction intents.
\done{The reasoned profiles and the multimodal descriptions of items, together with their ID embeddings, are propagated over the user-item interaction graph to absorb the high-order collaborative signals.}
On the other hand, we develop the threshold-controlled denoising and topology-aware enhancement strategies to refine the suboptimal item-item graph, \done{which are applied to both the multimodal and ID item representations to improve the accuracy of item representation learning.}
Extensive experiments on three publicly available datasets demonstrate that \acs{MLLMRec} achieves the state-of-the-art performance.
The source code is provided at \href{https://github.com/Yuzhuo-Dang/MLLMRec.git}{https://github.com/Yuzhuo-Dang/MLLMRec.git}.

\end{abstract}

%%
%% The code below is generated by the tool at http://dl.acm.org/ccs.cfm.
%% Please copy and paste the code instead of the example below.
%%

%\ccsdesc[500]{Computing methodologies~Neural networks}

%%
%% Keywords. The author(s) should pick words that accurately describe
%% the work being presented. Separate the keywords with commas.
%% A "teaser" image appears between the author and affiliation
%% information and the body of the document, and typically spans the
%% page.
%\begin{teaserfigure}
%  \includegraphics[width=\textwidth]{figures/sampleteaser}
%  \caption{Seattle Mariners at Spring Training, 2010.}
%  \Description{Enjoying the baseball game from the third-base
%  seats. Ichiro Suzuki preparing to bat.}
%  \label{fig:teaser}
%\end{teaserfigure}

%\received{20 February 2007}
%\received[revised]{12 March 2009}
%\received[accepted]{5 June 2009}

%%
%% This command processes the author and affiliation and title
%% information and builds the first part of the formatted document.
\maketitle

%!TEX root = ./0Manuscript.tex
\section{Introduction} \label{Introduction}
%111111
With the exponential growth of data scale and the diversification of content forms on online platforms, \ac{MMRS} have emerged as a pivotal technology for mitigating the information overload~\citep{DBLP:conf/mm/MuZT0T22, DBLP:conf/www/WeiHXZ23}.
By integrating the various item modalities (e.g., text, image, and audio) with the historical user-item interactions, \ac{MMRS} facilitate a more nuanced understanding of user preferences, thereby delivering highly personalized services in increasingly complex information environments~\citep{DBLP:journals/tmm/WangWYWSN23, DBLP:conf/mm/LiLW0NK24}.

The evolution of \ac{MMRS} has progressed through several distinct paradigms.
Pioneering research, exemplified by VBPR~\citep{DBLP:conf/aaai/HeM16}, focuses on simple feature fusion by augmenting the ID embeddings with the visual features.
Subsequently, to capture the high-order collaborative signals, graph-based methods like MMGCN~\citep{DBLP:conf/mm/WeiWN0HC19} and GRCN~\citep{DBLP:conf/mm/WeiWN0C20} introduce \ac{GCNs} to propagate the multimodal information across the user-item interaction graph.
Recognizing that the item-side correlations are often underutilized, some studies shift their focus toward mining the latent structural patterns among items.
For instance, LATTICE~\citep{DBLP:conf/mm/Zhang00WWW21} constructs the tunable item-item graphs by computing the cosine similarities between items based on the raw and projected modal features.
Further, FREEDOM~\citep{DBLP:conf/mm/ZhouS23} argues that the frozen item-item graphs constructed using only the raw features yield superior performance.
Most recently, the burgeoning field of \ac{LLMs} open new frontiers for \ac{MMRS}~\citep{DBLP:conf/recsys/BaoZZWF023, DBLP:conf/naacl/WangJCYZCFLHY24, DUAN2025114509}.
For example, LLMRec~\citep{DBLP:conf/wsdm/WeiRTWSCWYH24} leverages the vast world knowledge of \ac{LLMs} to generate the supplementary attributes for users and items.
Moreover, DOGE~\citep{DBLP:conf/aaai/MengMJLW25} employs \ac{MLLMs} to address the issue of granularity inconsistency across modalities, achieving the cross-modal relation mining.

\begin{figure}[t]
	\centering
	\subfigbottomskip=-1pt
	\subfigcapskip=-1pt
	\subfigure{
		\includegraphics[width=0.965\columnwidth]{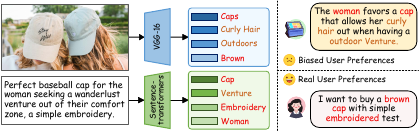}
		\label{1a}
	}
	\\ {\small (a) Noise interferes	with user preference learning}
	\vspace{3mm}
	\\
	\begin{minipage}{0.409\columnwidth}
		\centering
		\subfigure{
		\includegraphics[width=1.\columnwidth]{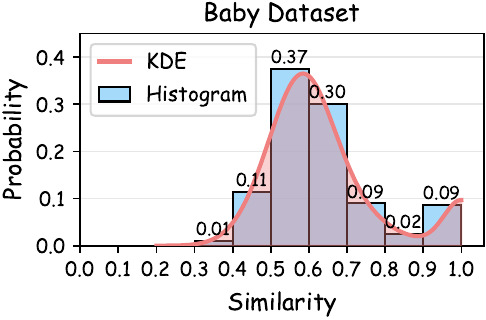}
		\label{1b}
		}
		\\ {\small (b) Similarity distribution}
	\end{minipage}
	\hspace{2mm}
	\begin{minipage}{0.522\columnwidth}
		\centering
		\subfigure{
			\includegraphics[width=1.\columnwidth]{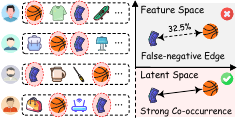}
			\label{1c}
		}
		\\ {\small (c) Missing co-occurrence}
	\end{minipage}
	\vspace{-1mm}
	\caption{Illustrations of (a) Noise interferes with user preference learning, (b) Distribution of similarity on the Baby dataset, and (c) Missing co-occurrence.} %
	\label{intro}
	\vspace{-3mm}
\end{figure}

Despite these remarkable achievements, existing \ac{MMRS} still exhibit the following two challenges in the representation learning at both the user and item sides:
(1) \textbf{Inaccurate initialization of user representations.}
Current approaches typically initialize the user representations through either random initialization or aggregating the modal features of the historically interacted items.
However, both paradigms are flawed.
Specifically, the former method is inherently behavior-agnostic, neglecting to leverage the rich preference signals in the interactions and thus exacerbating the cold-start issue.
The latter approach is often noise-contaminated, as it fails to filter out the modal features encoded by pre-trained models, leading to the indiscriminate aggregation of irrelevant features (e.g., the brown-marked noise in Figure~\ref{1a}).
More critically, these methods are incapable of capturing the latent interaction intents hidden in the user decision-making process.
Consequently, they are forced to rely on the sophisticated subsequent behavioral modeling to compensate for the informational deficiencies in the initial representations.
(2) \textbf{Structural noise in the item-item graph topology.}
The widely used $K$NN sparsification strategy~\citep{DBLP:journals/jmlr/ChenFS09} for the item-item graph construction is limited by its fixed number of neighbors.
This mechanism compels items lacking strongly similar neighbors to associate with low-similarity ones, inevitably introducing false-positive edges.
As the evidence reveals in Figure~\ref{1b}, approximately 12\% of edges in the Baby dataset link items with a similarity score below 0.5.
In addition, the $K$NN strategy depends exclusively on the similarity of modal features while ignoring the latent audience co-occurrence correlations, which are the vital collaborative signals in the field of recommender systems~\citep{liu2023co}.
Consequently, items that are semantically distant but behaviorally correlated remain disconnected (e.g., the basketball and genouillere in Figure~\ref{1c}), resulting in false-negative edges.
Both types of structural noise disrupt the efficacy of the subsequent graph convolution, ultimately leading to semantic drift in the item representation learning.

To address such limitations, we propose a novel \ac{MLLMRec} for multimodal recommendation.
Specifically, to ensure high-quality initialization of user representations, we first utilize the MLLM to convert the item images into semantic descriptions, thereby facilitating the cross-modal alignment.
Then, these visual descriptions are integrated with the textual metadata to form the comprehensive multimodal descriptions of items.
Distinct from conventional feature extraction, we embed these item-level semantic descriptions into the historical interaction sequences to construct the natural language descriptions of the user behaviors.
Building upon this, we leverage the advanced cognitive priors of the MLLM to reason over these descriptions, distilling latent interaction intents while effectively filtering modality-inherent noise to yield the behavior-aware and robust initial user preference profiles.
\done{These profiles, together with the multimodal descriptions of items and their ID embeddings, are then propagated over the user-item interaction graph to absorb the high-order collaborative signals.}
In addition, to rectify the structural noise in the item-item graph, we design two refinement strategies.
A threshold-controlled denoising strategy is implemented to prune the spurious edges with low similarity, while a topology-aware enhancement strategy is developed to recover the latent audience co-occurrence correlations.
Together, these strategies construct a more reliable item-item graph topology, providing a robust structural foundation for the subsequent graph convolution to capture high-order \done{features from both the multimodal and ID item representations} with greater precision.
\done{Finally, the multimodal and ID representations are concatenated to form the final user and item embeddings, and an InfoNCE contrastive loss is employed to align the two views.}
Extensive experiments on three datasets including Baby, Sports, and Clothing validate the effectiveness and robustness of \ac{MLLMRec}.

Our main contributions can be summarized as follows:
\begin{itemize}[leftmargin=*]
	\item We propose an innovative multimodal recommendation paradigm that utilizes the MLLM to reason about the user preference profiles, thereby distilling the latent interaction intents and filtering out the modality-inherent noise. \done{Moreover, these reasoned representations are propagated together with the ID embeddings over the user-item graph and aligned via an InfoNCE contrastive loss, yielding robust final user and item embeddings.}

	\item We develop two plug-and-play graph refinement strategies, comprising threshold-controlled denoising and topology-aware enhancement, to eliminate the low-similarity semantic edges and recover the latent co-occurrence links, respectively, thereby constructing the more reliable item-item graph.

	\item Extensive experiments conducted on three publicly available datasets demonstrate that \ac{MLLMRec} \done{achieves the state-of-the-art performance on most evaluation metrics, and the ablation studies verify the effectiveness of each module.}
	
\end{itemize}

%!TEX root = ./0Manuscript.tex
\section{Related Works}
\label{Related Works}
\subsection{Multimodal Recommendation}
Unlike conventional recommender systems that rely solely on the user-item interactions, \ac{MMRS} leverage the rich item-side modalities to facilitate a more comprehensive understanding of user preferences~\citep{DBLP:conf/sigir/Yi0OM22, DBLP:journals/kbs/DangPZCCC25}.
This field has evolved through several key paradigms.
Early efforts, such as VBPR~\citep{DBLP:conf/aaai/HeM16}, introduce the visual features as auxiliary signals to bolster the \ac{CF}.
With the advent of \ac{GCNs}, researchers shift toward the high-order representation learning.
For instance, MMGCN~\citep{DBLP:conf/mm/WeiWN0HC19} performs message passing on the modality-specific user-item interaction graphs.
To further capture the latent item-side correlations, subsequent works like LATTICE~\citep{DBLP:conf/mm/Zhang00WWW21} and FREEDOM~\citep{DBLP:conf/mm/ZhouS23} construct the item-item semantic graphs to perform \ac{GCNs}.
However, these methods ignore the different impacts of various modalities on the user decisions.
To address this, MGCN~\citep{DBLP:conf/mm/Yu0LB23} incorporates the behavior-aware modal weights, while GUME~\citep{DBLP:conf/cikm/LinMWLZ024} explicitly models user interests through the extended modality interactions.
More recently, \ac{SSL} has been introduced into \ac{MMRS} to mitigate the data sparsity and the modal noise.
For example, MMSSL~\citep{DBLP:conf/www/WeiHXZ23} employs data augmentation and contrastive learning to mine the inter-modal dependencies.
Moreover, MENTOR~\citep{DBLP:conf/aaai/00030000N25} argues that the excessive modality alignment may lead to the loss of interaction information, proposing a multi-level alignment strategy instead.
ModalSync~\citep{DBLP:conf/www/Liu0ZZB25} further integrates the supervised and self-supervised paradigms, achieving the synergistic optimization of the behavior modeling and multimodal learning.
However, these methods primarily focus on the item feature extraction while resorting to the shallow and incorrect initializations for the user representations.
Furthermore, the constructed item-item graph is noisy and incomplete, which impede the learning of robust item representations.

\subsection{LLMs-Enhanced Recommendation}
With the superior capabilities demonstrated by \ac{LLMs} in natural language processing tasks, researchers have increasingly explored their potential applications in the recommender systems~\citep{DBLP:journals/corr/abs-2311-02089, DBLP:conf/aaai/WangCOWHSGXZCLZ24, DBLP:journals/tois/LinDXLCZLWLZGYTZ25}.
Early explorations, such as P5~\citep{DBLP:conf/recsys/Geng0FGZ22}, directly employs a series of prompts to convert various recommendation metadata into a generative sequence-to-sequence framework.
Chat-Rec~\citep{DBLP:journals/corr/abs-2303-14524} feeds the user behavior sequences into Chat-GPT to generate the personalized recommendations.
Subsequently, to align the vast world knowledge of LLMs with the recommendation task, TALLRec~\citep{DBLP:conf/recsys/BaoZZWF023} uses LoRA~\citep{DBLP:conf/iclr/HuSWALWWC22} for the parameter-efficient fine-tuning of LLaMA, thereby improving its applicability to recommendation.
Moreover, ED$^2$~\citep{DBLP:conf/www/0005ZL00HZ0SD0Z25} proposes a dual dynamic indexing mechanism that enables the synergistic fusion between semantic knowledge and historical interactions.
Moreover, to overcome the inherent knowledge limitations of \ac{LLMs}, RecMind~\citep{DBLP:conf/naacl/WangJCYZCFLHY24} integrates the database querying and web searching to improve the reasoning accuracy, while introducing the self-inspiring learning for better planning.
Furthermore, CIKGRec~\citep{DBLP:conf/aaai/HuLJNDC0R25} designs a cross-domain contrastive learning framework that aligns the auxiliary information domain with the recommendation task.
Recently, several studies pivot toward \ac{MLLMs} to handle richer content formats.
For instance, DOGE~\citep{DBLP:conf/aaai/MengMJLW25} employs \ac{MLLMs} to translate the item images into the descriptive text, effectively unifying disparate the modal granularities.
Meanwhile, NEGGEN~\cite{ji2025generating} leverages the generative prowess of \ac{MLLMs} to synthesize the modality balanced negative samples, thereby providing the robust supervisory signals.
However, the existing MLLMs-based methods primarily treat \ac{MLLMs} as the sophisticated feature translators or data augmentors.
They overlook the potential of \ac{MLLMs} to perform preference reasoning by synthesizing the multimodal contents and the interaction patterns.

%!TEX root = ./0Manuscript.tex
\section{Task Formulation}
%\section{Methodology} \label{Approach}
Let $\mathcal{U} = \{u\}$ and $\mathcal{I} = \{i\}$ denote the sets of users and items, respectively.
Their observed historical interactions are represented by a Boolean matrix $\mathbf{R} = [r_{u,i}] \in \{0,1\}^{|\mathcal{U}| \times |\mathcal{I}| }$, where $r_{u,i}=1$ indicates an implicit feedback between user $u$ and item $i$, and $r_{u,i}=0$ otherwise.
Beyond the collaborative backbone, each item $i \in {\mathcal I}$ is associated with the multimodal attributes ${\mathcal M} = \{v,t\}$, where $v$ and $t$ denote the visual and textual domains, respectively~\citep{DBLP:conf/mm/ZhouS23, DBLP:conf/cikm/LinMWLZ024}.
In addition, we represent the collection of item images as $ \mathcal{D}^v = \{i^v\} $ and the corresponding textual descriptions as $ \mathcal{D}^t = \{i^t\} $.

Given the interaction matrix $\mathbf{R}$ and the multimodal item metadata $\{{\mathcal D}^v, {\mathcal D}^t\}$, the objective of our work is to learn a predictive function ${\mathcal F}: ({\mathbf R}, \mathcal{D}^v, \mathcal{D}^t) \to {\hat y_{ui}} $, where $\hat{y}_{ui} \in [0, 1]$ denotes the predicted preference score of user $u$ toward item $i$. The framework $\mathcal{F}$ aims to capture the synergistic effects between collaborative signals and multimodal semantics.
Finally, items are ranked in descending order of their predicted scores $\hat{y}_{ui}$ to generate a personalized top-$K$ recommendation list for each user.
%!TEX root = ./0Manuscript.tex
\section{Methodology} \label{Approach}
\begin{figure*}[t]
	\centering
	\includegraphics[width=\textwidth]{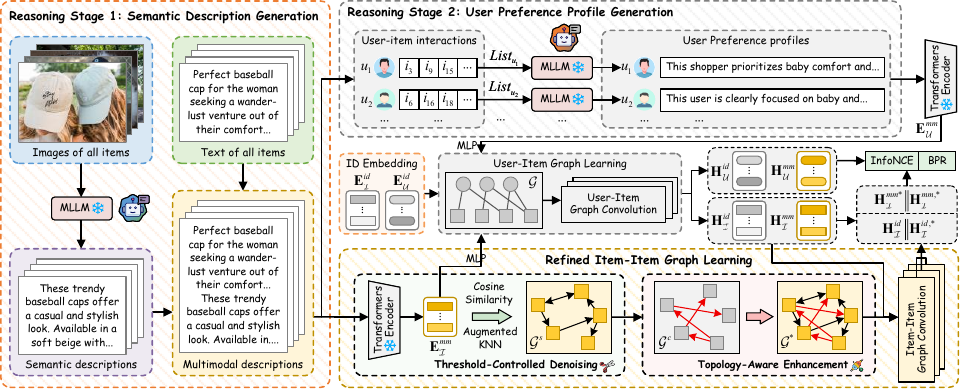}
	\vspace{-6mm}
	\caption{The overall framework of \ac{MLLMRec}. First, the MLLM transforms the images into the semantic descriptions, which are fused with the text to yield the multimodal descriptions. Next, \ac{MLLMRec} constructs the behavioral description lists for users, which are fed into the MLLM to reason about the user preference profiles. \done{Then, the reasoned profiles and the multimodal descriptions, together with the ID embeddings, are propagated over the user-item interaction graph to absorb the collaborative signals.} Meanwhile, \ac{MLLMRec} refines the item-item graph through two designed strategies, \done{which are applied to both the multimodal and ID item representations. Finally, the collaborative and multimodal representations are concatenated and aligned via the InfoNCE loss, and optimized by the BPR loss.}} %
	
	\label{framework}
	\vspace{-1mm}
\end{figure*}

Figure~\ref{framework} illustrates the overall framework of our proposed \ac{MLLMRec}, which synergizes the reasoning stages\done{, the collaborative representation learning, and} the refined item-item graph learning for multimodal recommendation.
In the subsequent sections, we elaborate on the design of each component.

\subsection{Reasoning Strategy}
To bridge the semantic gap across heterogeneous modalities and homogenize the data granularities, we first convert the item images into semantic descriptions.
Subsequently, we construct the personalized prompts for each user to guide the MLLM in reasoning about the complex behavioral patterns from the user-item historical interactions, thereby distilling the purified user preference profiles that contain the latent interaction intents.

\subsubsection{Semantic Description Generation}
Leveraging the sophisticated cross-modal understanding of \ac{MLLMs}, \ac{MLLMRec} generates the high-quality semantic descriptions for item images, which is formulated as follows:
\begin{equation}
	i^s = \text{MLLM} \left( \text{Prompt}_1, i^v \right),
\end{equation}
where $ i^s \in {\mathcal D}^s$ denotes the generated semantic description of image $i^v \in {\mathcal D}^v$, and the specific template of Prompt$_1$ is as follows:
\begin{tcolorbox}[colframe = black!70!white,
	colback = black!6!white,
	colbacktitle = black!55!white,
	left = 3.5mm,
	right = 3.5mm,
	top = 1.8mm, bottom = 1.6mm, boxsep = -0.5mm, toptitle = 1.5mm, bottomtitle = 1.5mm,
	title={Prompt$_1$: Semantic Description Generation}]
	
	\{\textit{image}\}: Please convert the given image into an accurate and concise textual description relevant to the \{dataset name\}, focusing on extracting key attributes that can influence the buying behavior of users, such as color, material, style, functionality, etc. To generate the textual description using a one-paragraph natural language overview in no more than 100 words.
	\tcblower
	\textbf{Input:} \{\textit{An Image}\}
	\\
	\textbf{Output:} \{A paragraph of semantic descriptions for this image\} %是否要换更亮的颜色
\end{tcolorbox}

This stage is pivotal for two reasons.
First, conventional visual encoders (e.g., VGG16~\cite{simonyan2014very}, ResNet~\cite{he2016deep}) often capture general features that may be tangential to the specific recommendation contexts.
By contrast, \ac{MLLMRec} designs a task-oriented prompt to guide the MLLM in selectively extracting the decision-relevant visual cues, such as aesthetic style.
This yields the semantic descriptions that are aligned with the recommendation task and filter out the noise of visual modality.
Second, this generative process projects the visual information into the linguistic space, effectively unifying the data formats of visual and textual modalities.
Such homogenization directly addresses the modality imbalance problem, which is a common bottleneck in the multimodal learning where models disproportionately rely on textual data due to the high noise-to-signal ratio and the semantic deficiency of raw visual features~\citep{ji2025generating, DBLP:conf/aaai/MengMJLW25}.
In summary, by transforming the images into the structured semantic description, we ensure that visual signals contribute equitably alongside textual signals to the preference learning.

Subsequently, we integrate $i^s$ with its raw textual metadata $i^t$ to construct a comprehensive multimodal description for each item.
The fusion process is formulated as follows:
\begin{equation}
	i^m = i^t \oplus i^s,
\end{equation}
where $i^m \in \mathcal{D}^m$ denotes the multimodal description of item $i$, and $\oplus$ represents the concatenation operation.
By integrating abstract visual descriptions with explicit textual metadata, facilitates the seamless fusion of multimodal information and effectively bridges the cross-modal semantic gap.
Consequently, it establishes a purified and enriched multimodal knowledge base that serves as a robust foundation for the subsequent module.

\subsubsection{User Preference Profile Generation} 
Distinguishing from conventional paradigms that rely on the behavior-agnostic random embeddings or the noise-contaminated feature aggregation, \ac{MLLMRec} capitalizes on the advanced cognitive reasoning of MLLMs to transform the raw interaction data into the purified user preference profiles. These profiles provide a high-fidelity semantic foundation for robust user representation learning.
Specifically, we first synthesize a behavioral descriptions list for each user by chronologically aggregating the multimodal descriptions of their historically interacted items.
This process is formulated as follows:
\begin{equation}
	\begin{array}{*{20}{c}}
		{List_u = \left[ {i^m} \right]},&{{r_{u,i}} = 1},
	\end{array}
\end{equation}
where $List_u$ denotes the behavioral description list of user $u$, and $i^m$ is the multimodal description of item $i$ that user $u$ has interacted with.
Subsequently, we embed $List_u$ into a meticulously designed prompt to guide the MLLM in reasoning about the user's profiles.
The generative process is formulated as:
\begin{equation}
	u^p = \text{MLLM} \left( \text{Prompt}_2, List_u \right),
\end{equation}
where $ u^p \in {\mathcal D}^p$ denotes the generated user preference profiles for user $u$, and the specific template of Prompt$_2$ is as follows:

\begin{tcolorbox}[colframe = black!70!white,
	colback = black!6!white,
	colbacktitle = black!55!white,
	left = 3.5mm,
	right = 3.5mm,
	top = 1.8mm, bottom = 1.6mm, boxsep = -0.5mm, toptitle = 1.5mm, bottomtitle = 1.5mm,
	title={Prompt$_2$: User Preference Profile Generation}]
	
	Please reason about the user preferences based on the following list of item descriptions that he or she has interacted with. The list is: \{\textit{behavioral description}\}. To generate the user preferences using a one-paragraph natural language in no more than 100 words.
	\tcblower
	\textbf{Input:} \{\textit{User's behavioral description list}\}
	\\
	\textbf{Output:} \{User's preference profiles in natural language\}
\end{tcolorbox}

By utilizing the extensive world knowledge inherent in MLLMs, the resulting user preference profile set ${{\mathcal D}^p} $ offers a more real user interests compared to the conventional paradigms.
It is noteworthy that since $List_u$ is derived from the interaction matrix $\mathbf{R}$, ${\mathcal D}^p$ already captures the latent motivations behind the user-item interactions.
\done{This inherent information richness provides a high-fidelity semantic foundation, which is further enhanced by the collaborative signals introduced in the subsequent learning stage.}

\subsection{Collaborative Representation Learning}
\done{To inject the collaborative signals into the preference reasoning paradigm, we design a dual-stream representation learning that jointly models the ID-based collaborative stream and the MLLM-driven multimodal stream, and then propagates both streams over the user-item interaction graph to obtain the high-order representations.
Specifically, we initialize the trainable ID embeddings for users and items, denoted as $\mathbf{E}_{\mathcal U}^{id} \in {\mathbb R}^{|\mathcal{U}| \times d}$ and $\mathbf{E}_{\mathcal I}^{id} \in {\mathbb R}^{|\mathcal{I}| \times d}$, respectively, via the Xavier initialization, which serve as the initial representations of the collaborative stream:}
\begin{equation}
	{\mathbf H}_{\mathcal U}^{id,(0)} = {\mathbf E}_{\mathcal U}^{id}, \quad {\mathbf H}_{\mathcal I}^{id,(0)} = {\mathbf E}_{\mathcal I}^{id}.
\end{equation}

\done{Meanwhile, for the multimodal stream, we first leverage the pre-trained text encoder to project the user preference profiles $\mathcal{D}^p$ and the item multimodal descriptions $\mathcal{D}^m$ into the latent space, which is formulated as follows:}
\begin{equation} \label{encoder}
	\mathbf{E}_{\mathcal U}^{mm} = \text{Encoder}\left( \mathcal{D}^p \right), \quad \mathbf{E}_{\mathcal I}^{mm} = \text{Encoder}\left( \mathcal{D}^m \right),
\end{equation}
\done{where $\mathbf{E}_{\mathcal U}^{mm} \in {\mathbb R}^{|\mathcal{U}| \times d_t}$ and $\mathbf{E}_{\mathcal I}^{mm} \in {\mathbb R}^{|\mathcal{I}| \times d_t}$ denote the extracted semantic features of users and items, respectively. To compress these high-dimensional features into a compact space, we further employ the \ac{MLPs} for the non-linear transformation, obtaining the initial representations of the multimodal stream.
Taking the user side as an example:}
\begin{equation}
	{\mathbf H}_{\mathcal U}^{mm,(0)} = \sigma \left( {\mathbf E}_{\mathcal U}^{mm} {\mathbf W}_1 + {\mathbf b}_1 \right) {\mathbf W}_2 + {\mathbf b}_2,
\end{equation}
\done{where ${\mathbf H}_{\mathcal U}^{mm,(0)} \in {\mathbb R}^{|\mathcal{U}| \times d}$ denotes the initial user representations of the multimodal stream, $\sigma(\cdot)$ is the LeakyReLU activation function. The weight matrices ${\mathbf W}_1 \in {\mathbb R}^{d_t \times d_1}$, ${\mathbf W}_2 \in {\mathbb R}^{d_1 \times d}$ and bias vectors ${\mathbf b_1} \in {\mathbb R}^{d_1}$, ${\mathbf b_2} \in {\mathbb R}^{d}$ are trainable parameters, where $d_1$ represents the dimension of the hidden layer, respectively.
Similarly, the item representations ${\mathbf H}_{\mathcal I}^{mm,(0)} \in {\mathbb R}^{|\mathcal{I}| \times d}$ are derived by applying an analogous projection to the item features ${\mathbf E}_{\mathcal I}^{mm}$.}

\done{Thereafter, we propagate both streams over the user-item interaction graph following the method in \cite{DBLP:conf/mm/ZhouS23} to capture the high-order collaborative signals. Formally, we construct the normalized adjacency matrix $\tilde{\mathbf A} = {\mathbf D}^{-\frac12} {\mathbf A} {\mathbf D}^{-\frac12}$, where ${\mathbf A} \in \{0,1\}^{(|\mathcal{U}|+|\mathcal{I}|) \times (|\mathcal{U}|+|\mathcal{I}|)}$ denotes the bipartite user-item adjacency matrix and ${\mathbf D}$ is the corresponding degree matrix. The graph convolution at the $l$-th layer is formulated as follows:}
\begin{equation}
	\left[{\mathbf H}_{\mathcal U}^{o,(l)} ; {\mathbf H}_{\mathcal I}^{o,(l)}\right] = \tilde{\mathbf A} \left[{\mathbf H}_{\mathcal U}^{o,(l-1)} ; {\mathbf H}_{\mathcal I}^{o,(l-1)}\right],
\end{equation}
\done{where $o \in \{id, mm\}$ indexes the two streams and $[\cdot\,;\,\cdot]$ denotes the vertical stacking operation. After $L_{ui}$ layers, we aggregate the multi-layer representations via summation as follows:}
\begin{equation}
	{\mathbf H}_{\mathcal U}^{o} = \sum\limits_{l = 0}^{L_{ui}} {\mathbf H}_{\mathcal U}^{o,(l)}, \quad
	{\mathbf H}_{\mathcal I}^{o} = \sum\limits_{l = 0}^{L_{ui}} {\mathbf H}_{\mathcal I}^{o,(l)}.
\end{equation}

\subsection{Refined Item-Item Graph Learning}
To mitigate the semantic drift in the item representation learning caused by the suboptimal item-item graph topology, we design two graph refinement strategies comprising (1) threshold-controlled denoising and (2) topology-aware enhancement.

\subsubsection{Threshold-Controlled Denoising}
Based on the extracted item multimodal features $\mathbf{E}_{\mathcal I}^{mm}$ obtained from Eq.~\eqref{encoder}, following the established practice~\cite{DBLP:conf/mm/ZhouS23}, we evaluate the semantic affinities between items by computing the cosine similarity of their multimodal features. This process is formulated as follows:
\begin{equation}
	{s}_{a,b} = \frac{{{{(\mathbf{e}_{{i_a}}^{mm})}^{\text{T}}}\mathbf{e}_{{i_b}}^{mm}}}{{\left\| {\mathbf{e}_{{i_a}}^{mm}} \right\| \cdot \left\| {\mathbf{e}_{{i_b}}^{mm}} \right\|}},
\end{equation}
%${\mathbf S} = [s_{a,b}]_{1 \leq a,b \leq |\mathcal{I}|} \in {\mathbb R}^{|\mathcal{I}| \times |\mathcal{I}|}$
where $s_{a,b}$ denotes the similarity score between item $a$ and item $b$, $\left\| \cdot \right\|$ represents the vector norm computation, and $(\cdot)^{\text{T}}$ is the matrix transpose operation.

While the conventional $ K $NN strategy~\citep{DBLP:journals/jmlr/ChenFS09} is widely adopted to sparsify the similarity matrix, it suffer from a rigid fixed-degree constraint.
This constraint forces items lacking the genuine semantic neighbors to establish the spurious connections with the low-similarity nodes.
As demonstrated by our empirical analysis in Figure~\ref{1b}, this phenomenon leads to an obvious presence of the false-positive edges (exceeding 10\% in the Baby dataset).
To rectify this, we introduce a threshold-controlled denoising strategy that combines the top-$K$ selection with a similarity constraint $\alpha$.
This ensures that only highly similar neighbors are retained for each item.
This strategy is formulated as follows:
\begin{equation}
	\tilde {s}_{a,b} = \left\{ {\begin{array}{*{20}{l}}
			{1,}&{{s}_{a,b} \in {\text{top-}}K_{s}({\mathbf s}_{a,:})} \ \ \text{and} \ \ {s}_{a,b} \geq \alpha,\\
			{0,}&{{\text{otherwise}},}
	\end{array}} \right.
\end{equation}
where vector ${\mathbf s}_{a,:} \in {\mathbb R}^{|\mathcal{I}|}$ represents the similarity scores between item $i_a$ and all items, the hyper-parameters $K_{s}$ and $ \alpha $ control the top-$K$ value and the pruning intensity, respectively.
The resulting matrix $ \tilde {\mathbf S} = [\tilde{s}_{a,b}] \in \{0,1\}^{|\mathcal{I}| \times |\mathcal{I}|}$ serves as the adjacency matrix for the denoised item-item semantic affinity graph ${\mathcal G^s} = \left\{ {\mathcal I}, {\mathcal E}^s \right\}$, where $\mathcal{I}$ is the set of item nodes and $\mathcal{E}^s = \{\left< i_a, i_b\right> | \tilde{s}_{a,b}=1\}$ denotes the edge set.
This formulation effectively filters out structural noise while preserving high-confidence semantic correlations.

\subsubsection{Topology-Aware Enhancement}
We argue that the semantic graph ${\mathcal G}^s$ constructed exclusively from the feature similarities fails to capture the item-item co-occurrence correlations that are essential collaborative signals embedded within the user-item interactions~\citep{liu2023co}.
To bridge this gap, we introduce a topology-aware enhancement strategy to recover these missing behavioral links.
Specifically, we quantify the audience co-occurrence between item pairs by computing the Jaccard similarity coefficient over their respective interacted user sets.
For any two items $i_a$ and $i_b$, their co-occurrence score $c_{a,b}$ is formulated as follows:
\begin{equation}
	c_{a,b} = J \left( \mathbf{r}_{:,a}, \mathbf{r}_{:,b} \right) = \frac{{\left| {\mathbf{r}_{:,a} \cap \mathbf{r}_{:,b}} \right|}}{{\left| {\mathbf{r}_{:,a} \cup \mathbf{r}_{:,b}} \right|}},
\end{equation}
where ${\mathbf r}_{:,a} \in {\mathbb R}^{|\mathcal{U}|}$ and ${\mathbf r}_{:,b} \in {\mathbb R}^{|\mathcal{U}|}$ are column vectors of the interaction matrix $ \mathbf{R} $, denoting the sets of users who have interacted with item $i_a$ and $i_b$, respectively.
The resulting matrix ${\mathbf C} = [c_{a,b}] \in {\mathbb R}^{|\mathcal{I}| \times |\mathcal{I}|}$ represents the audience co-occurrence matrix.
Given the sparsity of interaction data, most item pairs exhibit limited co-occurrence.
Consequently, to ensure structural robustness, we utilize a weighted $K$NN strategy to retain only the top-$K$ neighbors with the highest co-occurrence scores for each item, which is formulated as follows:
\begin{equation}
	\tilde {c}_{a,b} = \left\{ {\begin{array}{*{20}{l}}
			{{c}_{a,b},}&{{c}_{a,b} \in {\text{top-}}K_{c}({\mathbf c}_{a,:})},\\
			{0,}&{{\text{otherwise}},}
	\end{array}} \right.
\end{equation}
where vector ${\mathbf c}_{a,:} \in {\mathbb R}^{|\mathcal{I}|}$ represents the audience co-occurrence scores between item $i_a$ and all items, and hyper-parameter $K_{c}$ controls the density of co-occurrence connections.
This resulting weighted matrix $ \tilde {\mathbf C} = [\tilde{c}_{a,b}] \in {\mathbb R}^{|\mathcal{I}| \times |\mathcal{I}|}$ serves as the adjacency matrix for the item-item audience co-occurrence graph $\mathcal{G}^c = \{\mathcal{I}, \mathcal{E}^{c}\}$, where  $\mathcal{I}$ is the set of item nodes and ${\mathcal E}^{c} = \{  \left \langle i_a,i_b \right \rangle | \text{ } \tilde { c}_{a,b}>0 \} $ denotes the edge set with non-zero co-occurrence scores.

Finally, to establish a holistic representation of item-side relationships, we perform a graph fusion by integrating the semantic affinity graph with the audience co-occurrence graph, which is formulated as follows:
\begin{equation}
	{\mathbf S}^* = \tilde {\mathbf S} + \tilde {\mathbf C},
\end{equation}
where $ {\mathbf S}^* = [s^*_{a,b}] \in {\mathbb R}^{|\mathcal{I}| \times |\mathcal{I}|} $ serves as the adjacency matrix of the topology-enhanced item-item graph $\mathcal{G}^* = \{\mathcal{I}, \mathcal{E}^*\}$.
The fused edge set $\mathcal{E}^* = \mathcal{E}^s \cup \mathcal{E}^c$ effectively integrates the multimodal semantics with the collaborative signals.

%These two graph refinement strategies prune low-similarity links and recover missing behavioral correlations, both of which are prevalent issues in the vanilla multimodal item-item graph.

\subsubsection{Item-Item Graph Convolution}
Building upon the refined graph $\mathcal{G}^*$, we employ LightGCN~\citep{DBLP:conf/sigir/0001DWLZ020}, a simplified and effective variant of \ac{GCNs}, to perform the neighborhood aggregation on the item representations of \done{both streams} obtained from the user-item graph convolution.

This process enables \ac{MLLMRec} to capture the high-order correlations among items.
Specifically, the graph convolution operation at the $l\text{-th}$ layer is formulated as follows:
\begin{equation}
	{\mathbf H}_{\mathcal I}^{o,(l)} = \left( ({\mathbf N})^{-\frac{1}{2}} {\mathbf S}^* ({\mathbf N})^{-\frac{1}{2}} \right) {\mathbf H}_{\mathcal I}^{o,(l-1)},
\end{equation}
where $o \in \{id, mm\}$ \done{indexes the two streams}, ${\mathbf H}_{\mathcal I}^{o,(l)} \in {\mathbb R}^{|\mathcal{I}| \times d}$ represents the item representations of stream $o$ at the $l\text{-th}$ layer, ${\mathbf H}_{\mathcal I}^{o,(0)} = {\mathbf H}_{\mathcal I}^{o}$ \done{denotes the item representations output by the user-item graph convolution}, $\mathbf{N}$ denotes the degree matrix of graph $\mathcal{G}^*$ and $ {\mathbf N}_{aa} = \sum\nolimits_b {{{s}}_{a,b}^*}$.
After that, we aggregate the representations across all hidden layers through a summation function, which is formulated as follows:
\begin{equation}
	{\mathbf{H}}^{o,*}_{\mathcal I} = \sum\limits_{l = 0}^{L_{ii}} {\mathbf H}_{\mathcal I}^{o,(l)},
\end{equation}
where ${\mathbf{H}}^{o,*}_{\mathcal I} \in {\mathbb R}^{|\mathcal{I}| \times d}$ \done{denotes the refined high-order item representations of stream $o$}, and hyper-parameter $L_{ii}$ controls the depth of the graph convolution.

\subsection{Fusion and Optimization}

\done{To combine the complementary signals from the two streams, we concatenate the collaborative and multimodal features to form the final user and item embeddings, which is formulated as follows:}
\begin{equation}
	{\mathbf H}_{\mathcal U} = \left[ {\mathbf H}_{\mathcal U}^{id} \, \| \, \beta \cdot {\mathbf H}_{\mathcal U}^{mm} \right], \quad
	{\mathbf H}_{\mathcal I} = \left[ {\mathbf H}_{\mathcal I}^{id,*} \, \| \, \beta \cdot {\mathbf H}_{\mathcal I}^{mm,*} \right],
\end{equation}
\done{where ${\mathbf H}_{\mathcal U}, {\mathbf H}_{\mathcal I} \in {\mathbb R}^{2d}$ denote the final representations of users and items, respectively, $\|$ denotes the concatenation operation, and $ \beta $ is a hyper-parameter used to control the weight of the multimodal stream.
To further align the two views, we introduce an InfoNCE contrastive loss~\citep{DBLP:journals/jmlr/GutmannH10} between the collaborative and multimodal representations. Taking the user side as an example, the contrastive objective is formulated as follows:}
\begin{equation}
	{\mathcal L}_{cl}^{u} = - \frac{1}{|\mathcal{B}|} \sum\limits_{b \in \mathcal{B}} \log \frac{\exp\left( {\mathbf H}_{\mathcal U}^{id}[b] \cdot {\mathbf H}_{\mathcal U}^{mm}[b] / \tau \right)}{\sum\nolimits_{b' \in \mathcal{B}} \exp\left( {\mathbf H}_{\mathcal U}^{id}[b] \cdot {\mathbf H}_{\mathcal U}^{mm}[b'] / \tau \right)},
\end{equation}
\done{where $\mathcal{B}$ denotes the training batch, $\tau$ is the temperature hyper-parameter, and ${\mathbf H}_{\mathcal U}^{id}[b]$ and ${\mathbf H}_{\mathcal U}^{mm}[b]$ are the collaborative and multimodal representations of the $b$-th user, respectively. The item-side contrastive loss ${\mathcal L}_{cl}^{i}$ is defined analogously, and the overall contrastive loss is ${\mathcal L}_{cl} = {\mathcal L}_{cl}^{u} + {\mathcal L}_{cl}^{i}$.}
\done{Finally, we optimize the model parameters via the \ac{BPR} loss~\citep{DBLP:conf/uai/RendleFGS09}, which encourages the model to assign higher preference scores to the observed interactions than to unobserved ones:}
\begin{equation}
	{{\mathcal L}_{BPR}} =  - \sum\limits_{(u,{i_a},{i_b}) \in {\mathcal T}} {\log \left(\sigma\left( {{{\hat y}_{u{i_a}}} - {{\hat y}_{u{i_b}}}} \right) \right)},
\end{equation}
where ${\hat y}_{ui} = {\mathbf h}_u^{\text{T}} \cdot {\mathbf h}_i$ represents the predicted preference score for user $u$ on item $i$, and $\mathcal{T} = \{ (u, i_a, i_b) \mid r_{u,i_a}=1, r_{u,i_b}=0 \}$ denotes the set of training triplets, where $i_a$ is an observed item and $i_b$ is a negative sample randomly drawn from the unobserved set.
\done{The overall training loss is formulated as follows:}
\begin{equation}
	{\mathcal L} = {\mathcal L}_{BPR} + \lambda_{cl} {\mathcal L}_{cl},
\end{equation}
\done{where $\lambda_{cl}$ is the hyper-parameter balancing the contrastive loss.}
Ultimately, items are ranked in descending order of $\hat{y}_{u,i}$ to generate the top-$K$ recommendation list for each user, where $K$ denotes the length of the recommendation list.

%!TEX root = ./0Manuscript.tex
\section{Experiments} \label{Experiments}

\subsection{Experiment Setup}

\subsubsection{Datasets}
\begin{table}[t]
%	\footnotesize
	\centering
	\caption{Statistics of the experimental datasets.}
	\vspace{-4mm}
	\setlength{\tabcolsep}{2mm}{
		\begin{tabular}{ccccc}
			\toprule
			Datasets & \#Users & \#Items & \#Interactions & Density \\
			\midrule
			Baby  & 19,445 & 7,050 & 160,792 & 0.117\% \\
			Sports & 35,598 & 18,357 & 296,337 & 0.045\% \\
			Clothing & 39,387 & 23,033 & 278,677 & 0.031\%	\\
			\bottomrule
	\end{tabular}}%
	\label{Statistics of the experimental datasets}%
	\vspace{-2mm}
\end{table}%

Following~\citep{DBLP:conf/mm/ZhouS23, xu2025cohesion}, we conduct experiments on three widely used datasets from the Amazon product repository\footnote[1]{\url{http://jmcauley.ucsd.edu/data/amazon}}: (i) Baby, (ii) Sports and Outdoors, and (iii) Clothing, Shoes, and Jewelry (referred to as Baby, Sports, and Clothing, respectively).
These datasets cover diverse product categories, interaction sparsities, and scales of user and items, alongside rich multimodal metadata (i.e., item images and textual descriptions).
To ensure the reliability of collaborative signals, we apply the 5-core filtering strategy to filter out the users and items with fewer than five interactions.
The statistics of the datasets after filtering are summarized in Table~\ref{Statistics of the experimental datasets}.
For model training, each dataset is randomly partitioned into training, validation and testing sets with a ratio of 8:1:1.

\subsubsection{Evaluation Metrics}
We adopt two widely recognized top-$K$ recommendation metrics to quantify model performance: Recall (R@$K$) and Normalized Discounted Cumulative Gain (NDCG, N@$K$).
Specifically, Recall measures the fraction of the testing items successfully recommended, while NDCG accounts for ranking quality by assigning higher weights to relevant items positioned earlier in the list.
Consistent with prior works, we report the average scores across all users in the testing set under $K \in \{10, 20\}$.

\subsubsection{Baselines}
%representative and current SOTA 
%To comprehensively evaluate the superiority of \ac{MLLMRec}, we establish two types of baselines: (i) traditional recommender systems: BPR~\citep{DBLP:conf/uai/RendleFGS09} and LightGCN~\citep{DBLP:conf/sigir/0001DWLZ020}; (ii) multimodal recommender systems: VBPR~\citep{DBLP:conf/aaai/HeM16} is the pioneering work, LATTICE~\citep{DBLP:conf/mm/Zhang00WWW21} and FREEDOM~\citep{DBLP:conf/mm/ZhouS23} focus on the item-item graph learning, SLMRec~\citep{DBLP:journals/tmm/TaoLXWYHC23} introduces self-supervised learning, LGMRec~\citep{DBLP:conf/aaai/GuoL0WSR24} introduces the hyper-graph, SOIL~\citep{DBLP:conf/mm/Su0L0024} mines the second-order user-item interactions, GUME~\citep{DBLP:conf/cikm/LinMWLZ024} mines the fine-grained attributes, SMORE~\citep{DBLP:conf/wsdm/OngK25} denoises the modal features in the spectral space, DOGE~\citep{DBLP:conf/aaai/MengMJLW25} enhances the hyper-knowledge graph with \ac{LLMs}, and ModalSync~\citep{DBLP:conf/www/Liu0ZZB25} aligns the multimodal features with the user behaviors.

%
To demonstrate the superiority of \ac{MLLMRec}, we compare it against the following three categories of representative baselines: \ac{CF}-based recommenders, general multimodal recommenders, and MLLMs-based multimodal recommenders. A brief description of each baseline is provided below:

\vspace{1mm}
\noindent i) \ac{CF}-based recommenders:
\begin{itemize}[leftmargin=*]%,nolistsep
	\item \textbf{BPR}~\citep{DBLP:conf/uai/RendleFGS09} establishes a foundational Bayesian ranking framework that utilizes the pairwise comparison of items to capture the personalized user preferences from the implicit feedback.
	
	\item \textbf{LightGCN}~\citep{DBLP:conf/sigir/0001DWLZ020} simplifies the vanilla GCN by omitting the redundant non-linear activations and feature transformations.
\end{itemize}

\noindent ii) general multimodal recommenders:
\begin{itemize}[leftmargin=*]
	\item \textbf{VBPR}~\citep{DBLP:conf/aaai/HeM16} integrates the visual features from item images into the BPR model, enabling the capture of visual preferences.
	
	\item \textbf{LATTICE}~\citep{DBLP:conf/mm/Zhang00WWW21} constructs modality-aware graphs to mine the latent item-item semantic from multimodal features.
	
	\item \textbf{SLMRec}~\citep{DBLP:journals/tmm/TaoLXWYHC23} employs multimodal data augmentation and contrastive learning to capture the cross-modal patterns.
	
	\item \textbf{FREEDOM}~\citep{DBLP:conf/mm/ZhouS23} freezes the item-item graph to improve efficiency and incorporates a degree-sensitive pruning strategy to denoise the user-item interaction graph.
	
	\item \textbf{LGMRec}~\citep{DBLP:conf/aaai/GuoL0WSR24} combines local topological learning with a global hyper-graph module to capture the high-order dependencies.
	
	\item \textbf{GUME}~\citep{DBLP:conf/cikm/LinMWLZ024} utilizes mutual information maximization to align explicit interaction features with extended interest profiles.
	
	\item \textbf{ModalSync}~\citep{DBLP:conf/www/Liu0ZZB25} synchronizes multimodal feature extraction with user behaviors using a staged co-training strategy.
	
	\item \textbf{COHESION}~\citep{xu2025cohesion} integrates modality fusion and representation learning via a dual-stage fusion strategy and a composite GCN.
\end{itemize}

\noindent iii) MLLMs-based multimodal recommenders:
\begin{itemize}[leftmargin=*]
	\item \textbf{LLM-CF}~\citep{sun2024large} introduce \ac{LLMs} into collaborative filtering through a novel instruction-tuning method and in-context learning.
	
	\item \textbf{DOGE}~\citep{DBLP:conf/aaai/MengMJLW25} leverages \ac{MLLMs} to generate cross-modal representations by aligning the image and text information.
	
	\item \textbf{NEGGEN}~\citep{ji2025generating} uses \ac{MLLMs} to construct robust negative samples via tailored prompting, providing better contrastive signals.
	
\end{itemize}

\subsubsection{Implementation Details}

In the reasoning phase, we employ Qwen2.5vl-7b as our MLLM, accessed via the Ollama framework.
For fair comparisons with the existing multimodal baselines, we utilize Sentence-Transformer~\citep{DBLP:conf/emnlp/ReimersG19} as the text encoder.
%Xavier~\citep{DBLP:journals/jmlr/GlorotB10}, Adam~\citep{DBLP:journals/corr/KingmaB14}
Specifically, the training batch size is 2048 and the learning rate is 0.0001.
All trainable parameters are initialized via the Xavier method and optimized by the Adam optimizer.
We set the output dimension $d$ to 512, the hidden layer dimension $d_1$ to 2,048, the semantic neighbor count $K_{s}$ to 10, and the convolution depth $L_{ui}$ and $ L_{ii} $ to 2 and 1, respectively.
For the similarity threshold $\alpha$, we perform a grid search in $\{0.5, 0.6, 0.7, 0.8\}$, and the co-occurrence neighbor count $K_{c}$ is searched over $\{5, 10, 15, 20\}$.
The multimodal stream weight $\beta$ and the contrastive loss weight $\lambda_{cl}$ are tuned in $\{0.0001, 0.001, 0.01, 0.1\}$, and the temperature $\tau$ is set to 0.2.
In addition, to prevent over-fitting, the training process is terminated if R@20 on the validation set does not improve for 20 consecutive epochs, with a cap training limit of 1,000 epochs.
All experiments are implemented using PyTorch and evaluated on a 24GB NVIDIA RTX4090 GPU.

The optimal values of the dataset-specific hyper-parameters (i.e., $\alpha$, $K_{c}$, $\beta$, and $\lambda_{cl}$) are summarized in Table~\ref{optimal hyper-parameters}.

\begin{table}[t]
	\centering
	\caption{The optimal hyper-parameter settings of \ac{MLLMRec}.}
	\vspace{-4mm}
	\setlength{\tabcolsep}{4mm}{
		\begin{tabular}{ccccc}
			\toprule
			Dataset & $\alpha$ & $K_{c}$ & $\beta$ & $\lambda_{cl}$ \\
			\midrule
			Baby & 0.7 & 15 & 0.0001 & 0.01 \\
			Sports & 0.7 & 10 & 0.01 & 0.001 \\
			Clothing & 0.7 & 15 & 0.01 & 0.001 \\
			\bottomrule
	\end{tabular}}%
	\label{optimal hyper-parameters}%
	\vspace{-2mm}
\end{table}%

\subsection{Overall Performance}
\begin{table*}[h]
	\centering
	\renewcommand{\arraystretch}{1.05}
%	\footnotesize
	\caption{Overall performance of the baselines and \ac{MLLMRec} on three datasets, with the best and second-best results highlighted by bold and underline, respectively.}%
	\vspace{-4mm}
	\setlength{\tabcolsep}{1.5mm}{
		\begin{tabular}{l|cccc|cccc|cccc}
			\toprule
			\multicolumn{1}{c}{\multirow{2.5}{*}{Models}} & \multicolumn{4}{c}{Baby}      & \multicolumn{4}{c}{Sports}    & \multicolumn{4}{c}{Clothing} \\
			\cmidrule(lr){2-5} \cmidrule(lr){6-9} \cmidrule(lr){10-13}   \multicolumn{1}{c}{} & R@10  & R@20  & N@10  & \multicolumn{1}{c}{N@20} & R@10  & R@20  & N@10  & \multicolumn{1}{c}{N@20} & R@10  & R@20  & N@10  & N@20 \\
			\midrule
			BPR (UAI'09) & 0.0357 & 0.0575 & 0.0192 & 0.0249 & 0.0432 & 0.0653 & 0.0241 & 0.0298 & 0.0187 & 0.0279 & 0.0103 & 0.0126 \\
			LightGCN (SIGIR'20) & 0.0479 & 0.0754 & 0.0257 & 0.0328 & 0.0569 & 0.0864 & 0.0311 & 0.0387 & 0.0340 & 0.0526 & 0.0188 & 0.0236 \\
			\midrule
			VBPR (AAAI'16) & 0.0423 & 0.0663 & 0.0223 & 0.0284 & 0.0558 & 0.0856 & 0.0307 & 0.0384 & 0.0281 & 0.0415 & 0.0158 & 0.0192 \\
			LATTICE (MM'21) & 0.0547 & 0.0850 & 0.0292 & 0.0370 & 0.0620 & 0.0953 & 0.0335 & 0.0421 & 0.0492 & 0.0733 & 0.0268 & 0.0330 \\
			SLMRec (TMM'23) & 0.0540 & 0.0810 & 0.0285 & 0.0357 & 0.0676 & 0.1017 & 0.0374 & 0.0462 & 0.0452 & 0.0675 & 0.0247 & 0.0303 \\
			FREEDOM (MM'23) & 0.0627 & 0.0992 & 0.0330 & 0.0424 & 0.0717 & 0.1089 & 0.0385 & 0.0481 & 0.0629 & 0.0941 & 0.0341 & 0.0420 \\
			LGMRec (AAAI'24) & 0.0644 & 0.1002 & 0.0349 & 0.0440 & 0.0720 & 0.1068 & 0.0390 & 0.0480 & 0.0555 & 0.0828 & 0.0302 & 0.0371 \\
			GUME (CIKM'24) & 0.0673 & 0.1042 & 0.0365 & 0.0460 & 0.0778 & 0.1165 & 0.0427 & 0.0527 & 0.0703 & 0.1024 & 0.0384 & 0.0466 \\
			ModalSync (WWW'25) & 0.0693 & 0.1045 & 0.0373 & 0.0462 & \underline{0.0821} & \underline{0.1189} & \textbf{0.0457} & \underline{0.0551} & \underline{0.0738} & \underline{0.1058} & \underline{0.0406} & \underline{0.0488} \\
			COHESION (SIGIR'25) & 0.0680 & 0.1052 & 0.0354 & 0.0454 & 0.0752 & 0.1137 & 0.0409 & 0.0503 & 0.0665 & 0.0983 & 0.0358 & 0.0438 \\
			\midrule
			LLM-CF (CIKM'24) & 0.0593 & 0.0874 & 0.0312 & 0.0409 & 0.0698 & 0.1057 & 0.0376 & 0.0465 & 0.0596 & 0.0886 & 0.0329 & 0.0402 \\
			DOGE (AAAI'25) & \underline{0.0718} & \underline{0.1096} & \underline{0.0391} & \underline{0.0482} & 0.0792 & 0.1185 & 0.0435 & 0.0536 & 0.0712 & 0.1036 & 0.0389 & 0.0468 \\
			NEGGEN (MM'25) & 0.0701 & 0.1065 & 0.0342 & 0.0438 & 0.0763 & 0.1114 & 0.0411 & 0.0506 & 0.0654 & 0.0961 & 0.0350 & 0.0441 \\
			
			\midrule
			\ac{MLLMRec} (Ours) & \textbf{0.0722} & \textbf{0.1105} & \textbf{0.0396} & \textbf{0.0491} & \textbf{0.0823} & \textbf{0.1223} & \underline{0.0454} & \textbf{0.0575} & \textbf{0.0742} & \textbf{0.1075} & \textbf{0.0410} & \textbf{0.0493} \\
			
			\bottomrule
	\end{tabular}}%
	\label{Overall Performance}%
	\vspace{0mm}
\end{table*}%
Table~\ref{Overall Performance} presents the experimental results of the baselines and \ac{MLLMRec} on three datasets, with the following key findings:

\textbf{(1) Multimodal signals enhance the preference modeling.}
Integrating the modal representations substantially improves the performance of CF-based recommenders. A direct comparison between BPR and its modality extension VBPR reveals that the latter achieves superior results across all datasets. This confirms that relying solely on the collaborative signals is insufficient to capture the fine-grained user preferences due to the interaction sparsity. Multimodal information, such as visual cues, provides a critical semantic supplement, effectively enriching the representation space.

\textbf{(2) MLLMs promote the multimodal feature learning.}
Compared to the general multimodal recommenders, MLLMs-based methods exhibit competitive performance. For instance, DOGE achieves the best baseline results on the Baby dataset. This competitiveness is attributed to the exceptional cognitive reasoning and cross-modal alignment abilities of MLLMs, which facilitate a paradigm shift from the shallow feature extraction to the deep semantic understanding. By transforming the raw data into high-level descriptors, these models better capture the core features, leading to more precise matching in complex recommendation tasks.

\textbf{(3) \ac{MLLMRec} achieves the state-of-the-art performance.}
Specifically, \ac{MLLMRec} achieves the best performance on 11 of the 12 evaluation metrics across the three datasets, and the second-best result on the remaining one. These results demonstrate that jointly modeling the MLLM-reasoned multimodal semantics and the collaborative signals, together with the item-item graph refinement, effectively enhances the recommendation performance.

\textbf{(4) The collaborative and multimodal signals are complementary.}
Compared with the \ac{CF}-based baselines (e.g., BPR and LightGCN) that rely solely on the collaborative signals, and the MLLM-based baselines (e.g., DOGE and NEGGEN) that exploit the multimodal semantics, \ac{MLLMRec} integrates both streams through the user-item graph convolution and aligns them via the InfoNCE contrastive loss. The consistent gains over both categories of baselines demonstrate that the collaborative and multimodal signals are complementary, and their synergistic integration is beneficial for multimodal recommendation.

\subsection{Ablation Study}
\begin{table}[t]
	\vspace{0.8mm}
	\centering
	%	\fontsize{9}{10}\selectfont
	\caption{Performance of \ac{MLLMRec} and its variants in terms of R@20 and N@20 on three datasets.}
	\vspace{-4mm}
	\setlength{\tabcolsep}{1mm}{
		\begin{tabular}{c|cc|cc|cc}
			\toprule
			\multicolumn{1}{c}{\multirow{2.5}{*}{Model Variant}} & \multicolumn{2}{c}{Baby} & \multicolumn{2}{c}{Sports} & \multicolumn{2}{c}{Clothing} \\
			\cmidrule(lr){2-3} \cmidrule(lr){4-5}  \cmidrule(lr){6-7} \multicolumn{1}{c}{} & R@20  & \multicolumn{1}{c}{N@20} & R@20  & \multicolumn{1}{c}{N@20} & R@20  & N@20 \\
			\midrule
			\rowcolor{gray!18} MLLMRec & \textbf{0.1105} & \textbf{0.0491} & \textbf{0.1223} & \textbf{0.0575} & \textbf{0.1075} & \textbf{0.0493} \\
			$w/o$ RS & 0.0990 & 0.0438 & 0.1156 & 0.0524 & 0.1012 & 0.0459 \\
			$w/o$ GD & 0.1094 & 0.0487 & 0.1214 & 0.0550 & 0.1068 & 0.0486 \\
			$w/o$ TE & 0.1087 & 0.0480 & 0.1209 & 0.0546 & 0.1061 & 0.0477 \\
			$w/o$ $\text{GCN}_{\text{II}}$ & 0.1037 & 0.0465 & 0.1030 & 0.0464 & 0.0957 & 0.0432 \\
			$w/o$ CL & 0.0998 & 0.0442 & 0.1174 & 0.0531 & 0.1023 & 0.0464 \\
			\bottomrule
	\end{tabular}}%
	\label{Ablation Study}%
	\vspace{-3mm}
\end{table}%

To evaluate the performance contributions of each core component, we design the following five model variants: (1) $w/o$ RS, which removes the reasoning strategy (i.e., the multimodal stream) and retains only the collaborative stream, (2) $w/o$ GD, which excludes the threshold-controlled denoising strategy, (3) $w/o$ TE, which abandons the topology-aware enhancement strategy, (4) $w/o$ $\text{GCN}_{\text{II}}$, which removes the item-item graph convolution, and (5) $w/o$ CL, which removes the InfoNCE contrastive loss.
Table~\ref{Ablation Study} lists the experimental results, and it reveals the following three observations:

\textbf{(1) The absence of any core component leads to performance degradation.}
This confirms the necessity of the reasoning strategy, the two graph-refinement strategies, the item-item graph convolution, and the contrastive learning objective for the optimal performance of \ac{MLLMRec}.

\textbf{(2) The reasoning strategy and the item-item graph convolution are the two most critical components.}
Specifically, in terms of R@20, $w/o$ RS suffers the largest degradation on Baby, whereas $w/o$ $\text{GCN}_{\text{II}}$ incurs the largest drop on Sports and Clothing.
This validates the two pillars of \ac{MLLMRec}: the MLLM-reasoned preference profiles provide high-quality user representations, and the refined item-item graph captures the high-order relationships among items, both of which are indispensable for accurate preference modeling.

\textbf{(3) The contrastive loss brings consistent gains, while the graph-refinement strategies play a supplementary role.}
$w/o$ CL degrades the performance on all three datasets, demonstrating that aligning the collaborative and multimodal signals via the InfoNCE objective is beneficial.
In contrast, $w/o$ GD and $w/o$ TE incur relatively smaller but steady drops, indicating that the threshold-controlled denoising and topology-aware enhancement strategies serve as effective complements that refine the item-item graph and further stabilize the representation learning.

%!TEX root = ./0Manuscript.tex
\section{Conclusion and Future Work} \label{Conclusion}
In this paper, we propose \ac{MLLMRec}, a novel framework that redefines multimodal recommendation by \done{synergizing the cognitive reasoning of MLLMs, the collaborative signals,} and graph structural refinement.
Specifically, we introduce a pioneering paradigm that leverages \ac{MLLMs} to reason about high-fidelity user preference profiles, effectively capturing the latent interaction intents behind the user behaviors.
\done{These reasoned profiles, together with the multimodal descriptions of items and the ID embeddings, are propagated over the user-item interaction graph to absorb the high-order collaborative signals, and the two views are further aligned via an InfoNCE contrastive loss.}
Furthermore, to address the structural noise in the item-item graph, we design the threshold-controlled denoising and topology-aware enhancement strategies to refine the suboptimal item-item graph, \done{which are applied to both the multimodal and ID item representations.}
Extensive experiments across three \done{public} datasets demonstrate that \ac{MLLMRec} \done{achieves the state-of-the-art performance on most evaluation metrics.}

In the future, we aim to investigate temporally aware reasoning strategies to capture the evolving dynamics of user interests, while exploring the use of implicit reasoning tokens to bypass explicit textual steps, thereby improving the computational efficiency.

%%
%% The acknowledgments section is defined using the "acks" environment
%% (and NOT an unnumbered section). This ensures the proper
%% identification of the section in the article metadata, and the
%% consistent spelling of the heading.
\begin{acks}
To Robert, for the bagels and explaining CMYK and color spaces.
\end{acks}

%%
%% The next two lines define the bibliography style to be used, and
%% the bibliography file.
\bibliographystyle{ACM-Reference-Format}
\bibliography{reference}

%%
%% If your work has an appendix, this is the place to put it.
\appendix

\end{document}